\documentclass[%
aip,
rsi,
twocolumn,
amsmath,amssymb,
superscriptaddress,
 reprint,%
]{revtex4-1}

\usepackage[utf8]{inputenc}
\usepackage{amsmath,amssymb,amsfonts,bm}
\usepackage{graphicx,epstopdf}  
\usepackage{color}
\usepackage{extarrows}
\usepackage{enumitem}
\usepackage{empheq}
\usepackage{graphicx}
\usepackage{float}

\renewcommand{\d}{{\rm d}}

\newcommand{\w}{\omega}

\newcommand{\ti}{\tilde}

\newcommand{\B}{\mbox{\tiny B}}

\newcommand{\tS}{\mbox{\tiny S}}

\newcommand{\T}{\mbox{\tiny T}}

\newcommand{\SB}{\mbox{\tiny SB}}

\newcommand{\la}{\langle}
\newcommand{\ra}{\rangle}

\newcommand{\Sec}[1]{Sec.\,\ref{#1}}
\newcommand{\App}[1]{Appendix\,\ref{#1}}
\newcommand{\nl}{\nonumber \\}
\newcommand{\be}{\begin{equation}}
\newcommand{\ee}{ \end{equation}}
\newcommand{\bsube}{\begin{subequations}}
\newcommand{\esube}{\end{subequations}}
\newcommand{\Eq}[1]{Eq.\,(\ref{#1})}
\newcommand{\Eqs}[1]{Eqs.\,(\ref{#1})}
\newcommand{\Fig}[1]{Fig.\,\ref{#1}}

\newcommand{\RN}[1]{%
  \textup{\uppercase\expandafter{\romannumeral#1}}%
}

\newcommand{\greater}{\mbox{\tiny $>$}}
\newcommand{\lesser}{\mbox{\tiny $<$}}
\newcommand{\lgter}{\mbox{\tiny $\lessgtr$}}

\begin{document}

\title{
Dissipatons as generalized Brownian particles  for open quantum systems: Dissipaton--embedded quantum master equation
}

\author{Xiang Li}
\affiliation{CAS Key Laboratory of Precision and Intelligent Chemistry, University of Science and Technology of China, Hefei, Anhui 230026, China
}
\author{Yu Su}
\affiliation{Hefei National Research Center for Physical Sciences at the Microscale,  University of Science and Technology of China, Hefei, Anhui 230026, China}
\author{Zi-Hao Chen}
\affiliation{CAS Key Laboratory of Precision and Intelligent Chemistry, University of Science and Technology of China, Hefei, Anhui 230026, China
}
\author{Yao Wang}
\email{wy2010@ustc.edu.cn}
\affiliation{Hefei National Research Center for Physical Sciences at the Microscale,  University of Science and Technology of China, Hefei, Anhui 230026, China}
\author{Rui-Xue Xu}
\affiliation{CAS Key Laboratory of Precision and Intelligent Chemistry, University of Science and Technology of China, Hefei, Anhui 230026, China
}
\affiliation{Hefei National Research Center for Physical Sciences at the Microscale,  University of Science and Technology of China, Hefei, Anhui 230026, China}

\author{Xiao Zheng}
\email{xzheng@fudan.edu.cn}

\affiliation{Hefei National Research Center for Physical Sciences at the Microscale,  University of Science and Technology of China, Hefei, Anhui 230026, China}
\affiliation{Department of Chemistry, Fudan University, Shanghai 200433, China}

\author{YiJing Yan}
\email{yanyj@ustc.edu.cn}
\affiliation{Hefei National Research Center for Physical Sciences at the Microscale,  University of Science and Technology of China, Hefei, Anhui 230026, China}

\date{\today}
\begin{abstract}
Dissipaton theory had been proposed as an exact and nonperturbative approach to deal with open quantum system dynamics, where the influence of Gaussian environment is characterized by statistical quasi-particles named as dissipatons.
In this work,
we revisit the dissipaton equation of motion theory and establish 
 an equivalent dissipatons--embedded quantum master equation (DQME), which gives rise to
 dissipatons as generalized  Brownian particles. 
 As explained in this work,
the DQME supplies a direct approach to investigate the statistical characteristics of dissipatons and thus the physically supporting hybrid bath modes.
 Numerical demonstrations are carried out on the electron transfer model,  exhibiting the transient statistical properties of the solvation coordinate. 

\end{abstract}

\maketitle

\section{Introduction}
 Open quantum  systems are ubiquitous in various realms of physics.\cite{Wei21,Bre02,Yan05187,Lou73,
 Sli90,Van051037,Kli97,Ram98,
 Aka15056002,Muk81509,Yan885160,Yan91179,
Che964565,Tan939496,Tan943049,
Dor132746,Kun22015101}
The system--environment interactions may lead to exchange of energy, particle and information between the open system and its surrounding environment.
In literature, the environment is frequently known as baths or reservoirs.
When the influence of surrounding environments (baths/reservoirs) cannot be neglected, quantum dissipation theories can be employed to describe the reduced system dynamics.
Many quantum dissipation theories,
such as
the hierarchical equations of motion (HEOM) \cite{Tan89101,Tan906676,Tan06082001,Yan04216,Xu05041103,Xu07031107,Tan20020901} and the stochastic equation of motion (SEOM) \cite{Sha045053,Yan04216, Hsi18014104,Hsi18014104} theories, 
have been proposed, with focus mainly on the reduced system density operator, $\rho_{\tS}(t)\equiv{\rm tr}_{\B}\rho_{\T}(t)$.

However, it is well known that the reduced system dynamics alone, with relevant information encoded in $\rho_{\tS}(t)$,
is insufficient to deal with all experimental measurements.
The entangled system--and--environment dynamics are also crucially important
in the study of strongly correlated quantum impurity
systems.
\cite{Imr02,Hau08}
For example, the entangled system--environment 
correlation functions are closely related to such as
spectroscopy, \cite{Zha15024112,Du20034102,Che21244105}
transport \cite{Mei922512,Gru1624514,Wan20041102,Du212155,Wan22044102}
and thermodynamics \cite{Kir35300,Gon20154111,Gon20214115}
in quantum impurity systems.
These properties are usually considered to be beyond the scope of traditional treatments by quantum dissipation theories.
To address this issue, Yan  proposed the
dissipaton equation of motion (DEOM) in 2014,
which provides a statistical quasi-particle
picture for the influence of
environment that can be either bosonic or fermionic.
\cite{Yan14054105,Yan16110306,Wan22170901}
The DEOM recovers fully the HEOM for the dynamics of the primarily reduced system density operator.
Meanwhile, the underlying dissipaton algebra \cite{Yan14054105,Wan20041102} also makes the DEOM straightforward
to study the hybrid bath dynamics, polarization and nonlinear coupling
effects.\cite{Zha15024112,Xu151816,Wan22170901,Zha16204109,Yan16110306,Xu17395,Xu18114103,Su224554}

The DEOM \cite{Yan14054105} is not only an {``exact'' (cf.\,comments at the end of \Sec{prelude})} and nonperturbative approach to the  real--time evolution of open quantum system plus hybrid bath mode, but also serves as a
prototype for the development of other equations of motion within the same framework of dissipaton theory but in different scenarios. 
For example, 
to study the Helmholtz free energy change due
to the isotherm mixing of the system and environment, we proposed two independent approaches, the equilibrium $\lambda$-DEOM and imaginary--time DEOM.\cite{Gon20154111}  Nonequilibrium $\lambda(t)$-DEOM is also recently developed to compute the nonequilibrium work distributions in system--environment mixing processes. \cite{Gon22054109} 
To exactly treat the nonlinear environment couplings, we propose to incorporate the stochastic fields, which resolve just the nonlinear environment coupling terms, into the DEOM construction. 
The resultant stochastic-fields-dressed (SFD) total Hamiltonian contains only linear environment coupling terms. On the
basis of that, the SFD-DEOM was constructed. \cite{Che21174111} 
Besides,  we also propose the dissipaton thermofield theory  and  obtain the system--bath
entanglement theorem for nonequilibrium
correlation functions.\cite{Wan22044102} 
All these theoretical ingredients comprise the family of dissipaton theories.

Remarkably, in above mentioned dissipaton theories, quasi-particle descriptions for baths
can provide a unified treatment for the reduced system and hybridized environment dynamics.
This is right the point of  constructing an exact dissipaton theory, 
 which is inspired by HEOM formalism and adopts \emph{dissipatons}, etymologically derived from ther verb ``dissipate'' and the suffix ``-on'', as quasi-particles
associated with the Gaussian bath.\cite{Yan14054105,
Yan16110306,Zha18780,Wan20041102}
In this work,
we revisit the dissipaton equation of motion theory and establish 
 an equivalent dissipatons--embedded quantum master equation (DQME).
 In DQME, instead of a hierarchical structure, all the system--plus--dissipatons degrees of freedom are incorporated into a single dynamic
equation.
Specifically, we will demonstrate that dissipatons as  generalized Brownian quasi-particles, whose distribution obeys a Smoluchowski dynamics.
The DQME provides a direct approach to investigate the statistical characteristics of dissipatons and makes it convenient to obtain the hybrid bath modes dynamics.
Therefore, the DQME itself thus serves as  an important member of the family of dissipaton theories.

The remainder of this paper is organized as follows. In \Sec{thsec2}, we briefly review the basic onsets of dissipaton theories. DQME is constructed in \Sec{thsec3}.  
In \Sec{thsec4}, we detailedly discuss the statistical characteristics of dissipatons.
The numerical demonstration is  given in \Sec{num}, with the electron transfer model.
 We summarize this paper in \Sec{thsecsum}.
Throughout the paper we set $\hbar=1$ and $\beta=1/(k_BT)$, with $k_B$ the Boltzmann constant and $T$ the temperature.

\section{Onsets of dissipaton theory and DEOM}\label{thsec2}
\subsection{Prelude}\label{prelude}
Let us start from the basic system--plus--bath settings in the bosonic environment 
scenarios.
{{ For brevity}, 
we only consider the single dissipative--mode cases, where the system--bath coupling assumes a linear form of
$H_{\SB}=\hat Q\hat{F}$.
%
%
The total composite Hamiltonian tractable
within the dissipaton theory  has the generic form of
\begin{align}\label{HSB_boson}
 H_{\T}=H_{\tS}+H_{\tS\B}+h_{\B}.
\end{align}
Both the system Hamiltonian $H_{\tS}$
 and the dissipative system mode $\hat Q$ is arbitrary,
whereas the hybrid reservoir bath mode $\hat{F}$
assumes to be linear.
This together with noninteracting reservoir  model of $h_{\B}$
constitutes the Gauss--Wick's environment ansatz,
where the environmental influence is fully characterized by
the hybridization bath reservoir correlation functions,
$
\la\hat{F}_{\B}(t)\hat{F}_{\B}(0)\ra_{\B}
$.
Here, both $\hat{F}_{\B}(t)\equiv e^{ih_{\B}t}\hat{F}e^{-ih_{\B}t}$
and the equilibrium canonical ensemble average, $\la(\,\cdot\,)\ra_{\B}
\equiv {\rm tr}_{\B}[(\,\cdot\,)e^{-\beta h_{\B}}]/%
{\rm tr}_{\B}(e^{-\beta h_{\B}})$ 
are defined in the bare--bath subspace.
It can be related to the hybridize spectral density $J_{\B}(\w)$ via the 
fluctuation--dissipation theorem \cite{Wei21}
\begin{align}\label{FDT}
  \la\hat{F}_{\B}(t)\hat{F}_{\B}(0)\ra_{\B}
  =\frac{1}{\pi}
      \!\int^{\infty}_{-\infty}\!\!{\rm d}\omega\,
        \frac{e^{-i\omega t}J_{\B}(\omega)}{1-e^{-\beta\omega}}.
\end{align}
where
\begin{align}\label{JB}
  J_{\B}(\omega)\equiv\frac{1}{2}\!\int^{\infty}_{-\infty}\!\!{\rm d}t\,
    e^{i\omega t} \la[\hat{F}_{\B}(t),\hat{F}_{\B}(0)]\ra_{\B}.
\end{align}

The concept of \emph{dissipatons} originates from 
an exponential series expansion
of the hybridized bath correlation function,
\cite{Yan16110306}
\bsube
\be \label{FBt_corr}
\la\hat{F}_{\B}(t)\hat{F}_{\B}(0)\ra_{\B}
=\sum^K_{k=1}\eta_k e^{-\gamma_k t}.
\ee
\be \label{FBt_corr_rev}
\la\hat{F}_{\B}(0)\hat{F}_{\B}(t)\ra_{\B}
=\sum^{K}_{k=1}\eta_{k}^{\ast} e^{-\gamma_k^{\ast} t}\equiv\sum^{K}_{k=1} \eta_{\bar k}^{\ast} e^{-\gamma_k t}.
\ee
\esube
The  exponential series expansion
on $\la\hat{F}_{\B}(t)\hat{F}_{\B}(0)\ra_{\B}$ can be achieved by adopting a certain  sum--over--poles expression
for the Fourier integrand on the right--hand--side of Eq.(2),
followed by the Cauchy's contour integration,\cite{Hu10101106,Hu11244106,Din11164107,Din12224103,Zhe121129}
or  using the time--domain Prony fitting decomposition
scheme.\cite{Che22221102}
Together with the time--reversal relation $\la\hat{F}_{\B}(0)\hat{F}_{\B}(t)\ra_{\B}
=\la\hat{F}_{\B}(t)\hat{F}_{\B}(0)\ra_{\B}^{\ast}$ for $t\geq 0$.
The second equality of \Eq{FBt_corr_rev} is due to the fact that
the exponents in \Eq{FBt_corr} 
 must be either real or complex conjugate paired, and thus
we may determine $\bar k$ in the index set $ \{k=1,2,...,K\}$
by the pairwise equality $\gamma_{\bar k}=\gamma_{k}^{\ast}$.
This is a crucial property.
The exponential series expansion in \Eqs{FBt_corr} and (\ref{FBt_corr_rev}) inspired the idea of relating each exponential mode of correlation function to a statistical  quasi-particle, i.e., a \emph{dissipaton}.\cite{Yan14054105}

It is worth noting that the decomposition in \Eqs{FBt_corr} and (\ref{FBt_corr_rev})
can be obtained via alternative methods including the Fano spectrum decomposition, \cite{Cui19024110,Zha20064107}
the discrete Fourier series, \cite{Zho08034106}
the extended orthogonal polynomials expansions, \cite{Liu14134106,Tan15224112,Nak18012109,
Ike20204101,Lam193721} the time--domain Prony fitting scheme,\cite{Che22221102} and others. 
These methods have expanded the scope of application of HEOM/DEOM to the bath correlations in type of $\{t^{m}e^{-\gamma t}\}$ with $m$ a positive integer. 
The dissipaton theory can be extended to these scenarios and is numerically exact in these cases.
However, the quest of the most general and exact decomposition scheme is still a challenge.
}

\subsection{Dissipaton equation of motion (DEOM)}\label{thsec2B}
The dissipaton theory begins with
the \emph{dissipatons decomposition} that reproduces the correlation function in \Eqs{FBt_corr} and (\ref{FBt_corr_rev}).  
It decomposes $\hat{f}$ into a number of dissipaton operators $\{\hat{f}_k \}$, as
\begin{align}\label{hatFB_in_f}
 \hat F=\sum^K_{k=1}  \hat{f}_{k}.
\end{align}
In accordance with the dissipatons decomposition,  the dynamical variables in DEOM are
the dissipaton density operators
(DDOs), \cite{Yan14054105,Yan16110306,Zha18780}
\be \label{DDO}
  \rho^{(n)}_{\bf n}(t)\equiv \rho^{(n)}_{n_1\cdots n_K}(t)
\equiv {\rm tr}_{\B}\big[
  \big(\hat{f}_{K}^{n_K}\cdots\hat{f}_{1}^{n_1}\big)^{\circ}\rho_{\T}(t)
 \big].
\ee
Here, $n=n_1+\cdots+n_{K}$, with $n_k\geq 0$
for the bosonic dissipatons.
The product of dissipaton operators inside $(\cdots)^\circ$
is considered as \emph{irreducible},  which satisfies
$(\hat{f}_{k}\hat{f}_{j})^{\circ}
=(\hat{f}_{j}\hat{f}_{k})^{\circ}$
for bosonic dissipatons.
Each $n$--particles DDO, $\rho^{(n)}_{\bf n}(t)$, is associated with
an ordered set of indexes, ${\bf n}\equiv \{n_1\cdots n_K\}$.
Denote for later use also ${\bf n}^{\pm}_{k}$ that differs from ${\bf n}$ only
at the specified $\hat{f}_{k}$-disspaton occupation number
$n_{k}$ by $\pm 1$.
The reduced system density operator is a member of DDOs, 
$\rho_{\bf 0}^{(0)}(t)\equiv \rho_{0\cdots 0}^{(0)}(t)\equiv \rho_{\tS}(t)$.

For presenting the related dissipaton algebra later,
we adopt hereafter the following notations,
\be\label{DDO_action}
\begin{split}
  &\rho^{(n)}_{\bf n}(t;\hat A^{\times})\equiv
  {\rm tr}_{\B}\Big[\big(\hat{f}_{K}^{n_K}\cdots\hat{f}_{1}^{n_1}\big)^{\circ}
  \hat A^{\times}\rho_{\T}(t)\Big],
\\
 &\rho^{(n)}_{\bf n}(t;\hat A^{\lgter})\equiv
  {\rm tr}_{\B}\Big[\big(\hat{f}_{K}^{n_K}\cdots\hat{f}_{1}^{n_1}\big)^{\circ}
  \hat A^{\lgter}\rho_{\T}(t)\Big],
\end{split}
\ee
where $\hat{A}$ is an arbitrary operator,  and  $A^{\times}\equiv \hat A^{\greater}-\hat A^{\lesser}$ with
$
\hat A^{\greater}\rho_{\T}(t)\equiv \hat A\rho_{\T}(t)$ and 
$\hat A^{\lesser}\rho_{\T}(t)\equiv \rho_{\T}(t)\hat A
$.
The dissipaton theory comprises the following three basic ingredients.

\noindent (\emph{i}) \emph{Onset of dissipaton correlations}.
To reproduce \Eq{FBt_corr} and (\ref{FBt_corr_rev}), the dissipatons correlation functions read
\bsube\label{eq9}
\begin{align}\label{fx_corr}
  \la \hat{f}^{\B}_{k}(t)\hat{f}^{\B}_{k'}(0)\ra_{\B}=\delta_{k k'}\eta_{k} e^{-\gamma_{k}t},
\\ \label{fx_corr_nu_rev}
  \la \hat{f}^{\B}_{k'}(0)\hat{f}^{\B}_{k}(t)\ra_{\B}=\delta_{k k'} \eta_{\bar k}^{\ast} e^{-\gamma_{k}t},
\end{align}
\esube
with $\hat{f}^{\B}_{k}(t)\equiv e^{ih_{\B}t}\hat{f}_{k}e^{-ih_{\B}t}$.
Each forward--backward pair of dissipaton correlation functions
is specified by a single--exponent $\gamma_k$. 

\noindent(\emph{ii})
\emph{Onset of generalized diffusion equation}. The generalized diffusion equation of a dissipaton operator
reads \cite{Yan14054105,Yan16110306}
\begin{align}\label{gendiffu}
   {\rm tr}_{\B}[(\partial_t \hat{f}_k)_{\B}\rho_{\T}(t)]
   =-\gamma_k{\rm tr}_{\B}\big[\hat{f}_k\rho_{\T}(t)\big].
\end{align}
Together with
$(\partial_t \hat{f}_k)_{\B} =-i[\hat{f}_k, h_{\B}]$, we obtain\cite{Yan14054105}
\be\label{gendiff}
 \rho^{(n)}_{\bf n}(t;h_{\B}^{\times})=-i\bigg(\sum_k n_k \gamma_{k}\bigg)\rho^{(n)}_{\bf n}(t).
\ee
Equation (\ref{gendiff}) is the generalized diffusion equation in terms of DDOs.

\noindent(\emph{iii}) 
 \emph{Onset of generalized Wick's theorem}.
For the left and right actions of dissipaton operators, there exist 
\be\label{genwicks}
\begin{split}
\rho^{(n)}_{\bf n}(t;\hat{f}_k^{\greater})
&=\rho_{{\bf n}^{+}_{k}}^{(n+1)}(t)
 +n_{ k} \eta_{ k}\rho_{{\bf n}^{-}_{ k}}^{(n-1)}(t),
\\
\rho^{(n)}_{\bf n}(t;\hat{f}_k^{\lesser})
&=\rho_{{\bf n}^{+}_{k}}^{(n+1)}(t)
 +n_{ k} \eta^{\ast}_{ \bar k}\rho_{{\bf n}^{-}_{ k}}^{(n-1)}(t) .
\end{split}
\ee
This is known as the generalized Wick's theorem.

Based on the above three onsets,
by applying the Liouville-von Neumann equation,
$
 \dot\rho_{\T}(t)=-i[H_{\tS}+h_{\B}+H_{\SB}, \rho_{\T}(t)]
$
in \Eq{DDO},
followed by using \Eqs{gendiffu}--(\ref{genwicks}),
we obtain \cite{Yan14054105}
\begin{align}\label{DEOM}
 \dot\rho^{(n)}_{\bf n}=
& -\Big(i{\cal L}_{\tS}+\sum_k n_k \gamma_{k}\Big)\rho^{(n)}_{\bf n}
  -i\sum_{k}\hat Q^{\times}\rho^{(n+1)}_{{\bf n}_{k}^+}
\nl&
  -i\sum_{k}n_{k}\big(\eta_{k}\hat Q^{\greater}
   -\eta_{\bar k}^{\ast}\hat Q^{\lesser}\big)
  \rho^{(n-1)}_{{\bf n}_{k}^-}.
\end{align}
This is the DEOM for DDOs, where ${\cal L}_{\tS}\equiv H_{\tS}^{\times}$ and the parameters are given in \Eqs{FBt_corr} and (\ref{FBt_corr_rev}).
The resulting DEOM fully recovers the HEOM formalism.\cite{Tan89101,Tan906676,Tan06082001,Yan04216,Xu05041103,Xu07031107,Tan20020901} The latter is the time--derivative equivalent to Feynman--Vernon influence functional,\cite{Fey63118} with the primary focus on the reduced system density operator.

\subsection{Hybrid mode moments versus DDOs}
Based on the generalized Wick's theorem in \Eq{genwicks}, it is easy to relate the hybrid mode moments, $\{\la \hat F^{n}(t) \ra \}$, to DDOs, since 
\begin{align}\label{w1}
  \!\!\la\hat{F}^n\ra=\!\!\sum_{m=0}^{\lfloor n/2 \rfloor}\!\!
  {n \choose 2m}\cdot(2m-1)!!
  \la {\hat F}_{\B}^{ 2}\ra_{\B}^{m}
  \cdot \la( \hat F^{n-2m})^{\circ}\ra,
\end{align}
$\hat F=\sum_{k}\hat f_k$ and 
\begin{align}\label{FDDO}
  \la(\hat F^n)^{\circ}\ra &= \sideset{}{'}\sum_{\{n_k\}}
  \frac{n!}{n_1 !\cdots n_K ! } \la\big(\hat f_K^{n_K}\cdots \hat f_1^{n_1}\big)^{\circ}\ra\nl 
  &=\sideset{}{'} \sum_{\{n_k\}}  \frac{n!}{n_1 !\cdots n_K ! }{\rm tr_{\tS}}\rho^{(n)}_{\bf n},
\end{align}
with the summation being subject to $n_1\!+\cdots + n_K\! =\! n$. We thus complete \Eq{w1} in terms of DDOs; see \Eq{FB2} for $\la \hat{ F}_{\B}^2\ra_{\B}$. The above results can also be obtained via the path integral influence functional approach.\cite{Zhu12194106} 
As special cases, we have 
\be \label{1o}
\la\hat{F}(t) \ra= \sum_{k=1}^K{\rm tr}_{\tS}\rho^{(1)}_{{\bf 0}_{k}^{+}}(t),
\ee
and 
\begin{align}\label{2o}
\la\hat{F}^2(t) \ra&= \sum_{k=1}^{K}\eta_k
+\sum_{k,j=1}^{K}  {\rm tr}_{\tS}\rho^{(2)}_{{\bf 0}_{kj}^{++}}(t).
\end{align}

\section{Dissipatons--embedded quantum master equation (DQME)}\label{thsec3}
\begin{figure*}[t]
\includegraphics[width=1.6\columnwidth]{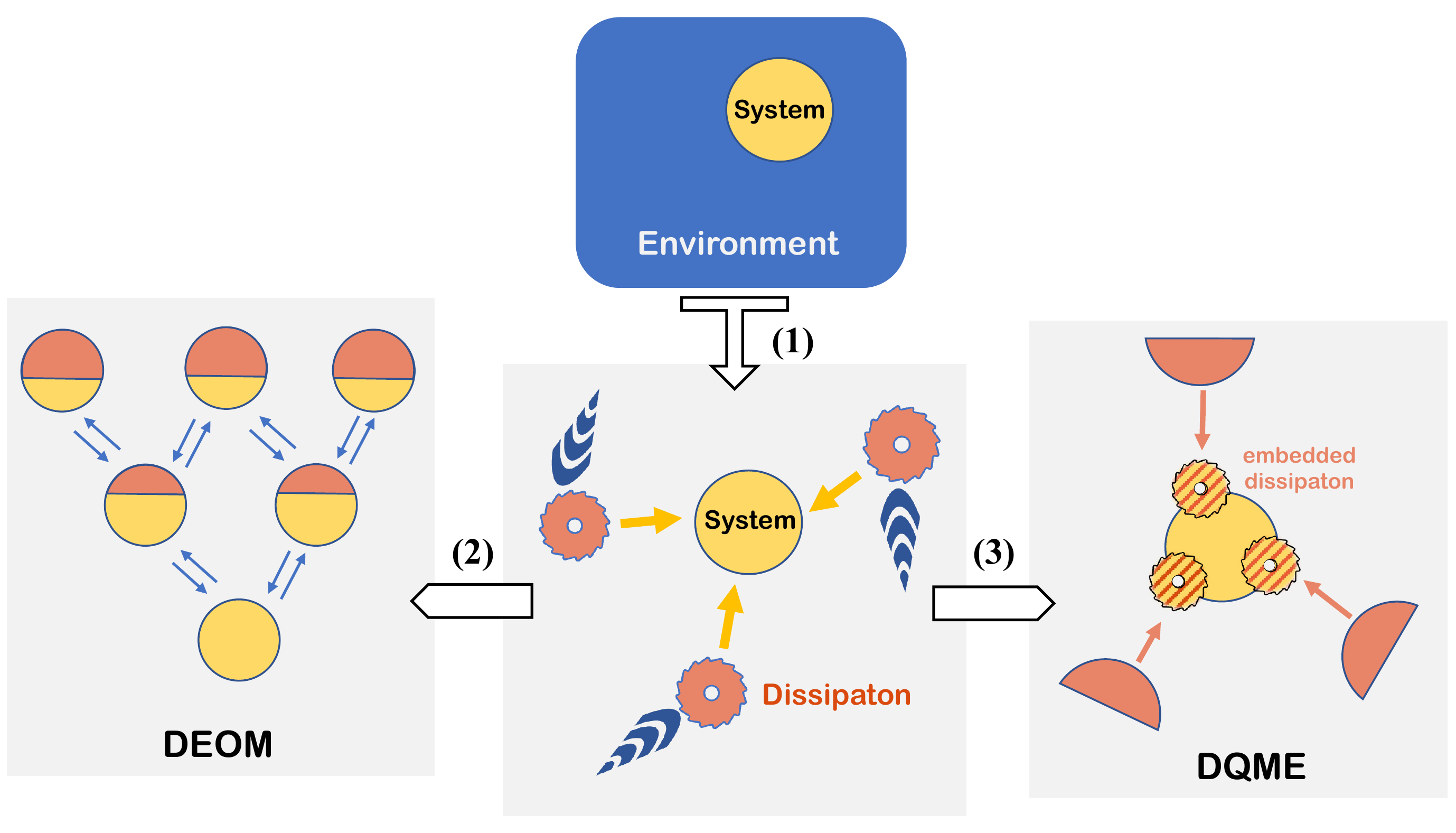}
\caption{A schematic illustration of dissipaton equation of motion (DEOM) versus dissipatons--embedded quantum master equation (DQME). Block arrow (1) is a mapping that decomposes  the  original environment (blue square) into multiple dissipatons (red gears), conserving the hybridized correlation function.
 Dissipatons behave as quantum Brownian particles.
%
Block arrow (2) represents the establishment of the DEOM of hierarchical structure. Each DDO (mixed red--yellow circle) incorporates a configuration of dissipatons. 
Block arrow (3) represents the establishment of  the DQME, where the disspaton degrees of freedom are embedded into the system (dashed red vortex).
Instead of a hierarchical structure, all the system--plus--dissipatons degrees of freedom are incorporated into a single dynamic
equation, influenced by dissipaton Smoluchowski operators (red arches).
}\label{fig1}
\end{figure*}
%

In this section, we will establish an equivalent dissipatons--embedded quantum master equation (DQME). This is concerned with the core system; i.e., the system--plus--dissipatons composite.
First of all, there is a one--to--one correspondence between the dissipaton operators and \emph{real} dimensionless variables: 
\be \label{map}
\hat f_{k}/\zeta_{k} \xleftrightarrow{\text{effective}} x_k \in (-\infty,\infty),
\ee
where
\begin{align} \label{coef}
\zeta_k\equiv \sqrt{(\eta_k+\eta^*_{\bar k})/2}
\ \ \text{and} \ \ 
\xi_k\equiv \frac{\eta_k -\eta^*_{\bar k}}{2i\zeta_k}.
\end{align}
The ``effective'' correspondence means that the statistical characteristics of hybrid modes [cf.\,\Eq{w1}] can be completely recovered by the moments of $\bm x\equiv \{x_k\}$; see \Eq{MF_2} and the discussions therein.
%

%
To proceed, we 
write the core--system distribution  as 
\be \label{13}
  \hat{\rho}({\bm x},t)
 = \sum_{\bf n} 
 \bigg[\rho_{\bf n}^{(n)}(t)
   \prod_{k} \phi_{n_k}(x_k)
 \bigg],
\ee
with the dissipaton subspace basis functions,
\begin{align}\label{eigen}
\phi_{n_k}(x_k)=(2\pi)^{-\frac{1}{4}} \frac{e^{-x^2_k /4}}{\zeta^{n_k}_k\sqrt{n_k! }}\phi_{n_k}^{\rm har}(x_k).
\end{align}
Involved  are the  standard harmonic oscillator wave functions,
\begin{align}
\phi^{\rm har}_n(x)=
(2\pi)^{-\frac{1}{4}}
(2^n n!)^{-\frac{1}{2}}
e^{-\frac{x^2}{4}} H_{n}\Big(\frac{x}{ \sqrt{2}}\Big),
\end{align} 
with $H_n(x)$ is the $n$th Hermite polynomial.
The reason for choosing $\{\phi_{n}(x)\}$ as the basis will be discussed in the next section; see  \Eq{MF_2} and  the discussions therein. 
Due to the orthogonal relation: \be 
\int\!\,e^{-x^2/2}H_{n}\Big(\frac{x}{ \sqrt 2}\Big)H_{n'}\Big(\frac{x}{ \sqrt 2}\Big){\d x}=(2\pi)^{\frac{1}{2}}2^{n}n!\delta_{nn'},
\ee
we can then recast \Eq{13} as
\begin{align} \label{rr}
  \rho^{(n)}_{\bf n}(t)=\!\!\int\hat{\rho}({\bm x},t)  \prod_k\bigg[ \Big(\frac{\zeta_k}{\sqrt 2}\Big)^{n_k} H_{n_k}\Big(\frac{x_k}{\sqrt 2}\Big)\bigg]  \d {\bm x}.
\end{align}
This is an alternative expression of DDOs, where $\d {\bm x}\equiv \d x_1 \cdots \d x_K$.

By using \Eq{13} with DEOM (\ref{DEOM}), followed by some detailed algebra in Appendix \ref{thApp1}, we obtain the DQME, 
\begin{align}\label{DQME}
\frac{\partial }{\partial t}\hat{\rho}({\bm x},t)=&-i[H_{\tS},\hat\rho]+\sum_k \hat{\Gamma}_k \hat\rho
-i\sum_k \zeta_k[\hat{Q},x_k \hat\rho]
\nl
&-\sum_k \xi_k \Big \{ \hat{Q},\frac{\partial \hat\rho}{\partial{x_k}} \Big \},
\end{align}
where
\be \label{gamma}
 \hat{\Gamma}_k \equiv \gamma_k \frac{\partial}{\partial{x_k}} \Big(\frac{\partial}{\partial{x_k}}+ x_k \Big ).
  \ee
This is the generalized Smoluchowski operator and
corresponds to the generalized diffusion equation (\ref{gendiffu}); see also \Eq{Ax}.
The last two terms in \Eq{DQME} correspond to the generalized Wick's theorem (\ref{genwicks}), describing the effect of system--bath coupling $H_{\tS\B}$; see also \Eqs{A7} and (\ref{A8}).  
It is worth reemphasizing that the parameters 
$\gamma_k$,  $\zeta_k$ and $\xi_k$ can all be complex, whereas the variable $x_k$ is real. 
As also known, 
$\gamma_{\bar k}=\gamma^{\ast}_{ k}$,
$\zeta_{\bar k}=\zeta^{\ast}_{k}$ 
and $\xi_{\bar k}=\xi^{\ast}_{k}$;
see \Eq{coef}, with $\bar{k}\in \{k=1,\cdots,K\}$.

Instead of a hierarchical structure, all the system--plus--dissipatons degrees of freedom are incorporated into a single dynamic
equation; see  \Eq{DQME},  \Fig{fig1} and the remarks therein. 
It is worth noting that the methodology here is
closely related to that in the  Fokker--Planck theory, Zusman theory and the pseudomode method. \cite{Tan06082001,Dal01053813,Tam18030402,Tam19090402,Che19123035,Lam193721,Hu907078}
The new DQME  recovers the reduced system dynamics as specified by $\rho_{\tS}(t)$ without any approximation.

\section{Statistical characteristics of dissipatons}\label{thsec4}
\subsection{Dissipatons as generalized Brownian particles}
In open quantum dynamics,  the wave-particle duality of both system and hybrid bath plays important roles.
In contrast to DEOM,  DQME describes the influence of hybrid bath in the aspect of waves,
with the picture of generalized Brownian particles for dissipatons.
%
To this end, we consider 
\begin{align}\label{rhoB}
  {P}(\bm x,t)\equiv {\rm tr}_{\tS}\hat{\rho}({\bm x},t),
\end{align}
and obtain 
\begin{align}\label{distri}
\frac{\partial }{\partial t}{P}(\bm x,t)=\sum_k\hat{\Gamma}_k {P}(\bm x,t) 
-\sum_k \frac{\partial}{\partial x_k} J_k(\bm x,t),
\end{align}
with the last term  the form of 
 ${\bm \nabla}\cdot{\bm J}$, arising from the system--bath coupling.
The  dissipaton probability current density vector  is given by
\be \label{Bt}
J_k(\bm x,t)=2 \xi_k  {\rm tr}_{\tS}\big[\hat{Q}\hat{\rho}({\bm x},t)\big].
\ee
Equation (\ref{distri}) provides a means to  the statistics of  hybrid bath. 
In particular, its equilibrium--state solution reads
\be \label{st}
{P}^{\rm st}={\hat \Gamma }^{-1} {\bm \nabla}\cdot{\bm J}^{\rm st}.
\ee
with $\hat \Gamma=\sum_k \hat\Gamma_k$.
However, to evaluate the current density, \Eq{Bt},  $\hat{\rho}({\bm x},t)$ via DQME (\ref{DQME}) is needed. 
From \Eq{st}, we can derive the input--output relations involving dissipaton moments and dissipative system mode; see Appendix \ref{B2} and comments therein. 
In the following, we will exploit the equivalent DEOM formalism to evaluate the hybrid bath statistics.
  
  \subsection{Transient dissipaton moments}
We will be interested in the expectation values of 
\begin{align}\label{Xn}
  {\bm x}_{\bf n}^{(n)}\equiv  x_{K}^{n_K}\cdots\ x_{1}^{n_1}.
\end{align}
To proceed, 
denote
\begin{align}\label{BigX}
\hat X^{(n)}_{\bf n}(t)&\equiv \int {\bm x}^{(n)}_{\bf n}\,\hat\rho({\bm x},t)\,{\rm d}{\bm x},
\end{align}  
and the transient dissipaton moments
\be \label{Mean}
\la  {\bm x}^{(n)}_{\bf n}(t)\ra={\rm tr}_{\tS}\hat X^{(n)}_{\bf n}(t) = \int {\bm x}^{(n)}_{\bf n}\,{P}({\bm x},t)\,{\rm d}{\bm x}.
\ee
%
On the other hand, we have (see \App{App2})
 \begin{align}\label{u1}
 \hat X^{(n)}_{\bf n}(t) =\sum_{\{m_k=0\}}^{\{\lfloor n_k/2 \rfloor\}} c_{\bf nm}  \rho^{(n-2m)}_{{\bf n} -2{\bf m}}(t),
 \end{align}
 %
 where
\be 
\rho^{(n-2m)}_{{\bf n} -2{\bf m}}(t)\equiv {\rm tr}_{\B}\big[
  \big(\hat{f}_{K}^{n_K-2m_K}\cdots\hat{f}_{1}^{n_1-2m_1}\big)^{\circ}\rho_{\T}(t)
 \big],
 \ee 
 and 
  \be\label{u1p}
 c_{\bf nm}=\prod_{k=1}^{K} \frac{1}{\zeta^{n_k- 2m_k }_k 2^{m_k}}\frac{n_k !}{m_k ! (n_k-2m_k)!}.
 \ee
%
We can recast \Eq{w1} as [cf.\,\Eq{w112} and discussion therein]
\begin{align}\label{MF_2}
\quad\la \hat{F}^n(t) \ra&=\!\int\! \Big(\sum_k \zeta_k x_k \Big)^n {{P}}(\bm x, t) \, \d {\bm x}
\nl
&= n! \sum_{\bf n} \Big( \prod_k \frac{\zeta^{n_k}_k}{n_k !}\Big) \la {\bm x}^{(n)}_{\bf n}(t)\ra.
\end{align} 
This also implies the one--to--one correspondence between the dissipaton operators and \emph{real} dimensionless variables, as specified in \Eq{map}.

\section{Numerical Demonstrations}\label{num}
For demonstration, we consider the electron transfer model,  \cite{Han0611438,Che07438}
\begin{align}\label{34}
\!\!\!H_{\tS}=(\epsilon+\lambda)|1\ra \la 1|+V(|1\ra\la 0|+|0\ra \la 1|),
\end{align}
with the dissipative system mode $\hat{Q}=-|1\ra\la1|$. Here, $|0\ra$ is the donor state and $|1\ra$ is the acceptor state, with  the
energy bias $\epsilon$, the interstate coupling $V$ and the solvent reorganization energy $\lambda$.
The latter arises from the bath spectral density, which takes the Brownian oscillator model, \cite{Yan05187}
\begin{align}
J_{\B}(\omega)=\frac{2\lambda \omega^2_0\zeta\omega}{(\omega^2-\omega_{0}^2)^2+\omega^2\zeta^2}.
\end{align}
Here in the scene of electron transfer, the hybrid mode $\hat{F}(t)$ is interpreted as ``solvation coordinate'' along the elecron transfer.  Its expectation $\la\hat{F}(t) \ra$ and standard deviation $\sigma_F={\la \hat{F}^2 (t)\ra^{1/2}}$  measure the reaction progress and the width of solvent wavepackage, respectively. 

The influence of anharmonic system induces the non-Gaussian dynamics of the environments. To measure  the non-Gaussianity of a distribution, there are two basic dimensionless quantities: \citep{Sha15} the skewness $K_3/\sigma^{3}_F$ and the kurtosis $K_4/\sigma^4_F$.  They characterize the asymmetry and the tailedness,  respectively, in terms of the $n$th cumulant,
\begin{align}
K_n(t)=\la \hat{F}^{n}(t)\ra-\sum^{n-1}_{m=1} {n-1 \choose m} K_{n-m}(t)\la \hat{F}^{m}(t)\ra,
\end{align}
with $K_1(t)=\la \hat{F}(t)\ra$. 

\begin{figure}[h]
\includegraphics[width=\columnwidth]{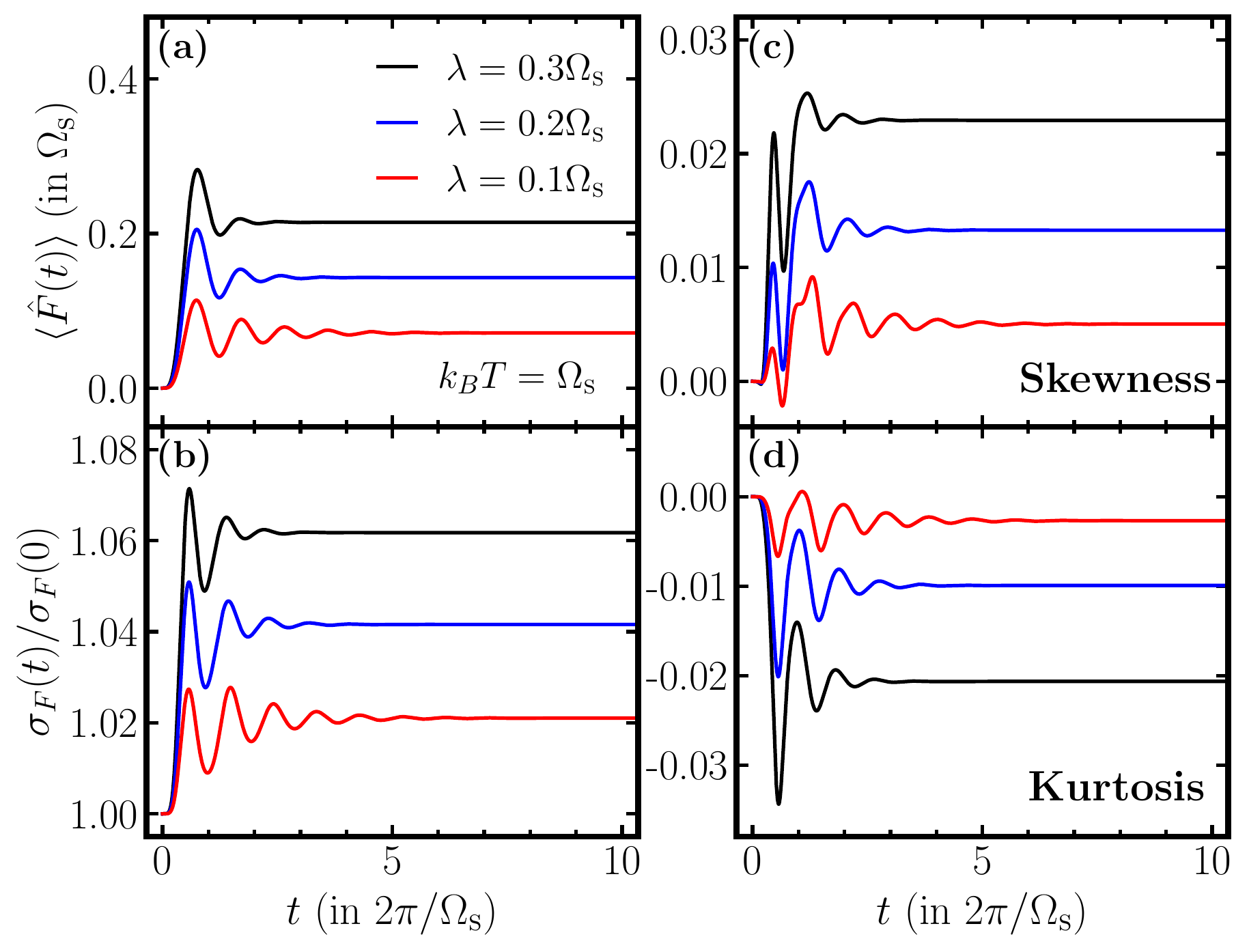}
\caption{Statistical characteristics of hybrid mode $\hat{F} (t)$ at three specified $\lambda$ and the given temperature, where $\Omega_{\tS}=\sqrt{\epsilon^2+4V^2}$: (a) The mean value; (b) The standard deviation $\sigma_F(t)/\sigma_F(0)$; (c) The skewness $K_3/\sigma^{3}_F$; (d) The kurtosis $K_4/\sigma^4_F$.  Other parameters are $
V = 0.4,\,
\omega_0 = 
\zeta = 1$ in unit of $\Omega_{\tS}$.
}
\label{fig2}
\end{figure}

Figure \ref{fig2} depicts the results of calculating the statistical characteristics  of transient hybrid mode $\hat{F}(t)$ at different coupling strength, i.e.  the reorganization energy $\lambda$.
 We can see that all the characteristics of $\hat{F}(t)$ oscillate at the frequency about $\Omega_{\tS}=\sqrt{\epsilon^2+4V^2}$,  the characteristic frequency  of system,  in short time evolution and converge to steady values in the long time asymptotic regime.  
 As the coupling strengthens,  the mean value and the standard deviation increase,  measuring the drifting and widening of distribution of reaction coordinate, respectively.   The skewness and kurtosis of hybrid mode deviate further from 0 as  coupling strengthens,  that indicates the influence of the anharmonic system violates the Gaussianity of the environment.
\begin{figure}[h]
\includegraphics[width=\columnwidth]
{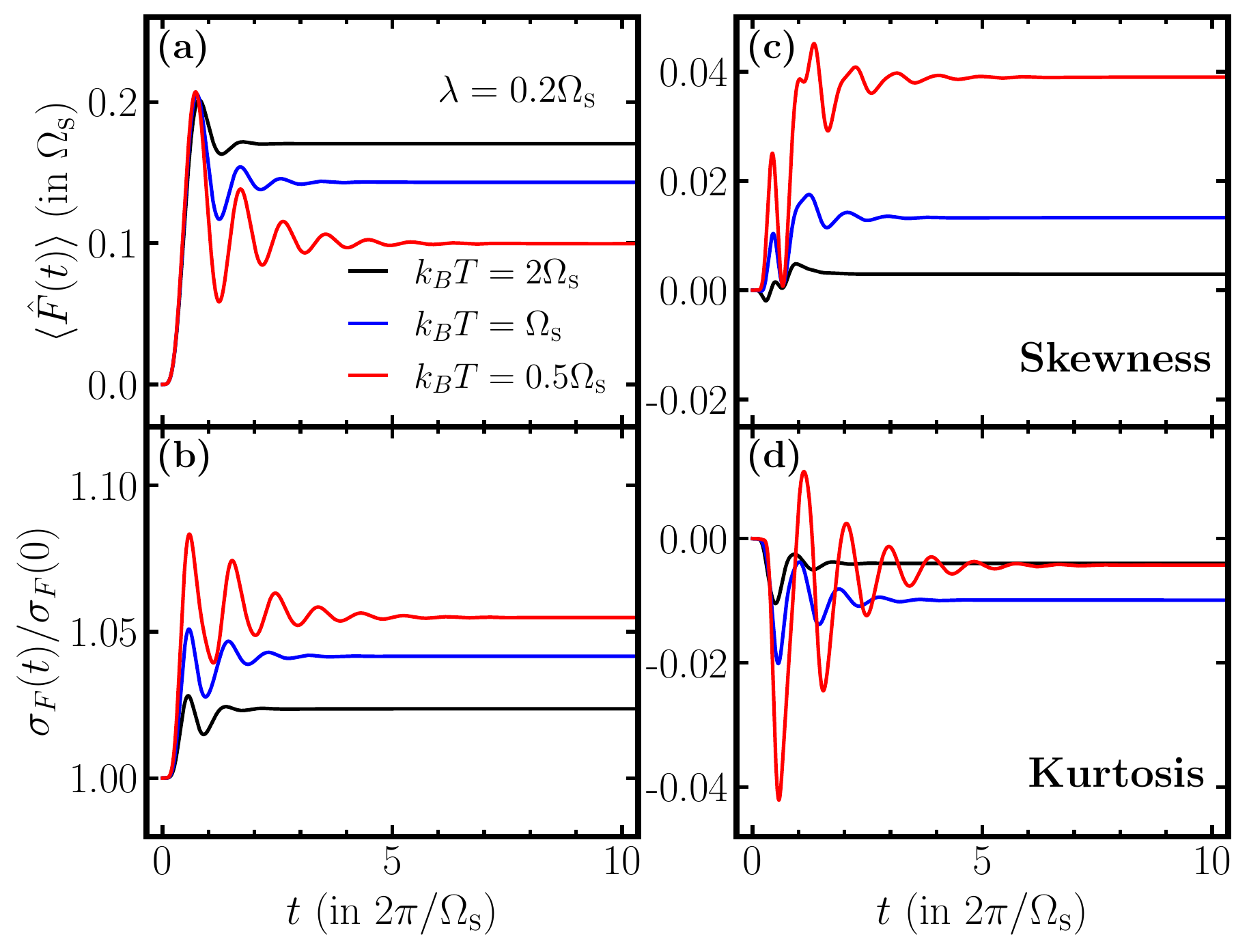}
\caption{Statistical characteristics of hybrid mode $\hat{F} (t)$ at three specified temperatures and the given $\lambda$, where $\Omega_{\tS}=\sqrt{\epsilon^2+4V^2}$: (a) The mean value; (b) The standard deviation $\sigma_F(t)/\sigma_F(0)$; (c) The skewness $K_3/\sigma^{3}_F$; (d) The kurtosis $K_4/\sigma^4_F$.  Other parameters are $
V = 0.4,\,
\omega_0 = 
\zeta = 1$ in unit of $\Omega_{\tS}$.
}
\label{fig3}
\end{figure}

 Figure \ref{fig3} shows the transient evolutions of statistical characteristics  of  the hybrid mode at different  temperatures.
Similar dynamical oscillation behaviors as that in \Fig{fig2} are displayed. On the other hand, when temperature increases,  $\la \hat{F} \ra$ at the steady state  increases (cf.\,\Fig{fig3}a),  while $\sigma_F(t)/\sigma_F(0)$ and the skewness reduce (cf.\,\Fig{fig3}b and 3c).  However,   the kurtosis does not varies monotunously in this regime, which suggests the complicated behaviors of the tailedness of relevant distributions.

\section{Concluding remarks}\label{thsecsum}

To summarize, in this work
we construct the dissipaton--embedded quantum master equation (DQME) from the DEOM theory via introducing the one--to--one correspondence between the dissipaton operators and \emph{real} dimensionless variables.
Instead of a hierarchical structure, all the system--plus--dissipatons degrees of freedom are incorporated into a single dynamic
equation in DQME. 
The new DQME recovers the reduced system dynamics as specified by $\rho_{\tS}(t)$ without any approximation. 
Moreover, the statistical characteristics of hybrid modes can be completely recovered.

The formalism of DQME reveals the picture of dissipatons as the Brownian quasi-particles interacting with the system. Based on the DQME,  we can discuss the evolution of dissipaton distribution under the influence of system and correlate the transient moments of dissipatons with DDOs.  

The fermionic DQME can also been readily established in a similar manner, which would benefit for the simulations on such as spintronic and superconductive systems.
The DQME formalism brings the possibility of achieving the quantum simulation of non-Markovian open quantum dynamics.
Since all the dissipaton degrees of freedom are represented by continuous real variables in DQME, this makes it a versatile formalism for incorporating matrix product states, real--space renormalization group, and other numerical methods. 
It is anticipated that DQME developed in this work would become an important tool for quantum mechanics of open systems.

\begin{acknowledgements}
Support from the Ministry of Science and Technology
of China (Grant No.\ 2021YFA1200103) and the National Natural Science Foundation of China (Grant Nos.\
21973086, 22103073 and 22173088) is gratefully acknowledged. 
\end{acknowledgements}

\appendix
\section{Derivation of DQME (\ref{DQME})}\label{thApp1}
For later use, we first rewrite \Eq{13} as
\be
  \hat{\rho}({\bm x},t)
 = \sum_{\bf n} 
 \bigg[\ti \rho_{\bf n}^{(n)}(t)
   \prod_{k} \ti \phi_{n_k}(x_k)
 \bigg],
\ee
with
\be 
\ti \rho_{\bf n}^{(n)}(t)= \rho_{\bf n}^{(n)}(t)\Big/\prod_k\zeta^{n_k}_k,
\ee
and
\begin{align}
\ti \phi_{n_k}(x_k)=(2\pi)^{-\frac{1}{4}} \frac{e^{-x^2_k /4}}{\sqrt{n_k! }}\phi_{n_k}^{\rm har}(x_k).
\end{align}
Note for $\ti\phi_{n}(x)$, we have \cite{Shi09164518}
\be 
x \ti{\phi}_n(x)
=\ti{\phi}_{n-1}(x)
   +(n+1) \ti{\phi}_{n+1}(x),
\ee
and
\be
  \frac{\partial \ti{\phi}_n(x)}{\partial {x} }
=-(n +1)\ti {\phi}_{n+1}(x).
\ee

According to DEOM (\ref{DEOM}), the EOM for $\ti \rho_{\bf n}^{(n)}(t)$ reads
\begin{align}\label{DEOM2}
\dot{\ti{\rho}}^{(n)}_{\bf{n}}(t)=&-\Big({i\cal{L}}_{\tS}+\sum_k {n_k}\gamma_k\Big) \ti{\rho}^{(n)}_{{\bf n}}(t)
\nl
&-i\sum_{k}\zeta_k \hat{Q}^{\times}\Big[\ti{\rho}^{(n+1)}_{{\bf n}^+_k}(t)+n_k\ti{\rho}^{(n-1)}_{{\bf n}^-_k}(t) \Big]
\nl
&+\sum_{k}  n_k \xi_{k}\hat{Q}^{\diamond}\ti{\rho}^{(n-1)}_{{\bf n}^-_k}(t),
\end{align}
where \Eq{coef} is used and  $\hat A^{\diamond}\rho\equiv \hat A\rho+\rho \hat A$.
With 
\begin{align}\label{Ax}
&\sum_{\bf n} \sum_k {n_k}\gamma_k \ti{\rho}^{(n)}_{{\bf n}} \prod_{j} \ti \phi_{n_{j}}(x_{j})
\nl
=&-\sum_k \gamma_k \frac{\partial}{\partial x_k} (x_k+\frac{\partial}{\partial x_k})\sum_{\bf n} \ti{\rho}^{(n)}_{\bf n} \prod_{j} \ti \phi_{n_{j}}(x_{j})
\nl
=&-\sum_k \gamma_k \frac{\partial}{\partial x_k} (x_k+\frac{\partial}{\partial x_k})\hat{\rho}({\bm x}),
\end{align}
\begin{align}\label{A7}
  &\sum_{\bf n} 
 \bigg[\Big(\ti \rho_{{\bf n}_{k}^{+}}^{(n+1)}+n_k\ti \rho_{{\bf n}_{k}^{-}}^{(n-1)}
   \Big)\prod_{j} \ti \phi_{n_{j}}(x_{j})
 \bigg]
\nl  
 =&\sum_{\bf n} 
 \bigg[\ti \rho_{{\bf n}}^{(n)} \prod_{j} \big[\ti \phi_{n_{j}-\delta_{kj}}(x_{j})+(n_k+1)\phi_{n_{j}+\delta_{kj}}(x_{j})\big]\bigg]
 \nl 
 = &x_k\hat{\rho}({\bm x}),
\end{align}
and
\begin{align}\label{A8}
&\sum_{\bf n}n_k\ti \rho_{{\bf n}_{k}^{-}}^{(n-1)}
   \prod_{j} \ti \phi_{n_{j}}(x_{j})
   \nl
   =&-\sum_{\bf n}\ti \rho_{{\bf n}_{k}^{-}}^{(n-1)}
\frac{\partial}{\partial x_k}\prod_{j} \ti \phi_{n_{j}-\delta_{kj}}(x_{j})
\nl
=&-\frac{\partial}{\partial x_k}\hat{\rho}({\bm x}),
\end{align}
we then obtain the DQME (\ref{DQME}) from \Eq{DEOM2} together with \Eqs{Ax}--(\ref{A8}).

\section{Moments of dissipatons }\label{App2}
\subsection{Basic relations and transient dissipaton moments}
Let us start with some basic relations:
\begin{align}\label{FB2}
  \la\hat F_{\B}^2\ra_{\B} =\sum_k\eta_k=\sum_k\eta_{\bar k}^{\ast} =\sum_k\zeta_k^2.
\end{align}
Applied here are \Eqs{FBt_corr}, (\ref{FBt_corr_rev}) and (\ref{coef}).
Moreover, 
\begin{align}\label{A10}
  {\bm x}_{\bf n}^{(n)} \frac{\partial}{\partial x_k}=
   \frac{\partial}{\partial x_k}{\bm x}_{\bf n}^{(n)}
  -n_k  {\bm x}_ {{\bf{n}}^-_k}^{(n-1)},
\end{align}
and [cf.\,\Eq{gamma}]
\begin{align}\label{A11}
  {\bm x}_{\bf n}^{(n)}\hat \Gamma_k &=
  \gamma_k \bigg[
  \Big(\frac{\partial^2}{\partial x^2_k}-n_k\Big){\bm x}_{\bf n}^{(n)}
  + 
  n_k(n_k -1){\bm x}_ {{\bf{n}}^{--}_{kk}}^{(n-2)}
 \nl
  &\quad
  - n_k \frac{\partial}{\partial x_k}{\bm x}_ {{\bf{n}}^-_k}^{(n-1)} +\frac{\partial}{\partial x_k} {\bm x}_ {{\bf{n}}^+_k}^{(n+1)}\bigg].
\end{align}

With the relation
\begin{align}
  \Big(\frac{x}{\sqrt 2}\Big)^n=\frac{n!}{2^n}\sum^{\lfloor n/2 \rfloor}_m \frac{1}{m!(n-2m)!} H_{n-2m}\Big(\frac{x}{\sqrt 2}\Big),
\end{align}
we can obtain $\hat X^{(n)}_{\bf n} $ in \Eq{BigX}
 \begin{align}
 \hat X^{(n)}_{\bf n}=\int \, {\bm x}^{(n)}_{\bf n} \rho(t;\{x_k \})\d {\bm x} 
=\!\!\sum^{\{[n_k/2]\}}_{\{m_k\}}\!\! c_{\bf mn} \rho^{(n-2m)}_{\bf{n} -2\bf{m}},
 \end{align}
 expressed by DDOs, where $ c_{\bf mn}$ is given in \Eq{u1p}.
  Inversely, we have
  \begin{align}\label{u2}
 \rho^{(n)}_{\bf n}(t) = \sum_{\{m_k=0\}}^{\{\lfloor n_k/2 \rfloor\}} \bar c_{\bf nm}  \hat X^{(n-2m)}_{\bf{n} -2\bf{m}}(t),
 \end{align}
 with 
  \begin{align}\label{u2p}
 \bar c_{\bf nm}=\prod_k \zeta^{n_k}_k 2^{-m_k} \frac{(-1)^{m_k}n_k !}{m_k ! (n_k-2m_k)!}.
 \end{align}
 
To show that \Eq{MF_2} is equivalent \Eq{w1}, 
we first rewrite \Eq{w1} as
\begin{align}\label{w112}
  \!\!\la\hat{F}^n\ra=&\sum_{m=0}^{\lfloor n/2 \rfloor}
  \frac{ n!}{2^m m! }
  \la {\hat F}_{\B}^{ 2}\ra_{\B}^{m}
\sideset{}{'} \sum_{\{n_k-2m_k\}}  \frac{1}{n_1 !\cdots n_K ! }{\rm tr_{\tS}}\rho^{(n-2m)}_{\bf n-2m}.
\end{align}
By further using the last identity in \Eq{FB2}, we then reproduce \Eq{w1}.
This validates  the effective mapping (\ref{map}), which conserve the statistics of the hybrid bath.

\subsection{Equilibrium dissipaton statistics}\label{B2}
	To obtain $\la {\bm x}_{\bf n}^{(n)} \ra$, the moments of $\{ x_k \}$ at thermal equilibrium, we first recast \Eq{st} as
\begin{align} 
\int {\bm x}_{\bf n}^{(n)} \,{\hat \Gamma}{P}^{\rm st}\,\d {\bm x} =\int {\bm x}_{\bf n}^{(n)} {\bm \nabla} \cdot {\bm J}^{\rm st}\,\d {\bm x} .
\end{align}
For \Eq{Xn}, \Eqs{A10} and (\ref{A11}) gives the input--output relation reading
\begin{align}\label{28}
\!\!\la {\bm x}^{(n)}_{\bf n}\ra\!=\!
\frac{2}{\gamma_{\bf n}^{(n)}}\! \sum_k\bigg[ \xi_kn_k \la \hat{Q} {\bm x}^{(n-1)}_{{\bf n}_{k}^{-}} \ra
+\gamma_k{n_k\choose 2}\la  {\bm x}^{(n-2)}_{{\bf n}_{kk}^{--}}\ra\bigg],
\end{align}
where $\gamma_{\bf n}^{(n)} \equiv \sum_k n_k\gamma_k$. 

To close the input--output relation, one may further consider for any operator $\hat A$, there exists the relation
\begin{align}
      &i\la [H_{\tS},\hat A] {\bm x}^{(n)}_{\bf n} \ra_{\rm st} 
     +i \sum_k \zeta_k\la [\hat{Q},\hat{A}]{\bm x}^{(n+1)}_{\bf n^+_k}\ra_{\rm st}
     \nl
      &+ \sum_k \xi_k n_k \la \{\hat A,\hat{Q}\} {\bm x}_{{\bf n}^-_{k}}^{(n-1)} \ra_{\rm st}
      \nl
      =&\sum_k \gamma_k\Big[ n_k \la \hat A {\bm x}_{\bf n}^{(n)}\ra_{\rm st}- n_k (n_k-1)\la \hat A{\bm x}_{{\bf n}^{--}_{kk}}^{(n-2)}\ra_{\rm st}\Big].
      \end{align}
It is evident that the recurrence relation  is closed when $\hat{A}$ belongs to a linear space $\mathbb{L}$ of operators.  Here, the linearly independent  basis set of $\mathbb{L}$ has two parts:   (\emph{i})  basis vectors $\hat V$ satisfying $[H_{\tS},\hat{V}]=0$,  $[\hat{Q},\hat{V}]=0$ and $\{\hat{Q},\hat{V}\}=0$;  (\emph{ii})   finite basis vectors  generated from the identical operator $\hat I$  via the operation $\{\hat Q,\cdot \, \}$,  $[\hat{Q},\cdot]$ and $[H_{\tS},\cdot\, ]$. As a result, the basis set is  $\{\hat I, \hat{Q},[H_{\tS},\hat{Q}],\hat{Q}^2,[\hat{Q}, [H_{\tS},\hat{Q}]],[H_{\tS}, [H_{\tS},\hat{Q}]],\{\hat{Q},[H_{\tS},\hat{Q}]\},  
\nl
\cdots\}$.\cite{Gu851310}

For example, for the spin--boson system where $H_{\tS}=\hat \sigma_{x}$, $\hat Q=\hat \sigma_{z}$, the linearly dependent basis of $\mathbb{L}$ is $\{\hat{I}, \hat{\sigma}_z, \hat{\sigma}_y, \hat{\sigma}_x\}$, since $[\hat \sigma_i,\hat \sigma_j]=2i\varepsilon_{ijk}\hat\sigma_{k}$ and $\{\hat \sigma_i,\hat \sigma_j\}=2\delta_{ij}\hat I$.
Then, when $\hat A=\hat\sigma_{x}$, $\hat \sigma_{y}$ and $\hat\sigma_{z}$, we  have
\begin{align}
&2\sum_k\zeta_k\la \hat\sigma_{y} {\bm x}_{{\bf n}^+_k}^{(n+1)}\ra 
\nl
=&\sum_k \gamma_k\Big[ n_k \la \hat\sigma_{x} {\bm x}_{\bf n}^{(n)}\ra
- n_k (n_k-1)\la \hat\sigma_{x}{\bm x}_{{\bf n}^{--}_{kk}}^{(n-2)}\ra\Big],
\end{align}
\begin{align}
&-2V\la \hat\sigma_{z} {\bm x}^{(n)}_{\bf n} \ra+2\sum_k\zeta_k\la \hat\sigma_{x} {\bm x}_{{\bf n}^+_k}^{(n+1)}\ra 
\nl
=&\sum_k \gamma_k\Big[ n_k \la \hat\sigma_{y} {\bm x}_{\bf n}^{(n)}\ra- n_k (n_k-1)\la \hat\sigma_{y} {\bm x}_{{\bf n}^{--}_{kk}}^{(n-2)}\ra\Big],
\end{align}
and
\begin{align}
&\quad -2V\la \hat \sigma_{y}{\bm x}^{(n)}_{\bf n} \ra+2\sum_k \xi_k n_k \la {\bm x}_{{\bf n}^-_{k}}^{(n-1)} \ra
\nl
&=\sum_k \gamma_k\Big[ n_k \la \hat \sigma_{z}{\bm x}_{\bf n}^{(n)}\ra- n_k (n_k-1)\la \hat \sigma_{z}{\bm x}_{{\bf n}^{--}_{kk}}^{(n-2)}\ra\Big].
\end{align}
separately.


\begin{thebibliography}{10}

\bibitem{Wei21}
U.~Weiss,
\newblock {\em Quantum Dissipative Systems},
\newblock World Scientific, Singapore, 2021,
\newblock 5$^{\rm th}$ ed.

\bibitem{Bre02}
H.~P. Breuer and F.~Petruccione,
\newblock {\em The Theory of Open Quantum Systems},
\newblock Oxford University Press, New York, 2002.

\bibitem{Yan05187}
Y.~J. Yan and R.~X. Xu, \newblock ``Quantum mechanics of dissipative systems,''
  Annu. Rev. Phys. Chem. {\bf 56}, 187 (2005).

\bibitem{Lou73}
W.~H. Louisell,
\newblock {\em Quantum Statistical Properties of Radiation},
\newblock Wiley, New York, 1973.

\bibitem{Sli90}
C.~P. Slichter,
\newblock {\em Principles of Magnetic Resonance},
\newblock Springer Verlag, New York, 1990.

\bibitem{Van051037}
L.~M.~K. Vandersypen and I.~L. Chuang, \newblock ``NMR techniques for quantum
  control and computation,'' Rev. Mod. Phys. {\bf 76}, 1037 (2005).

\bibitem{Kli97}
C.~F. Klingshirn,
\newblock {\em Semiconductor Optics},
\newblock Springer-Verlag, Heidelberg, 1997.

\bibitem{Ram98}
J.~Rammer,
\newblock {\em Quantum Transport Theory},
\newblock Perseus Books, Reading, Mass., 1998.

\bibitem{Aka15056002}
Y.~Akamatsu, \newblock ``Heavy quark master equations in the Lindblad form at
  high temperatures,'' Phys. Rev. D {\bf 91}, 056002 (2015).

\bibitem{Muk81509}
S.~Mukamel, \newblock ``Reduced equations of motion for collisionless molecular
  multiphoton processes,'' Adv. Chem. Phys. {\bf 47}, 509 (1981).

\bibitem{Yan885160}
Y.~J. Yan and S.~Mukamel, \newblock ``Electronic dephasing, vibrational
  relaxation, and solvent friction in molecular nonlinear optical lineshapes,''
  J. Chem. Phys. {\bf 89}, 5160 (1988).

\bibitem{Yan91179}
Y.~J. Yan and S.~Mukamel, \newblock ``Photon echoes of polyatomic molecules in
  condensed phases,'' J. Chem. Phys. {\bf 94}, 179 (1991).

\bibitem{Che964565}
V.~Chernyak and S.~Mukamel, \newblock ``Collective coordinates for nuclear
  spectral densities in energy transfer and femtosecond spectroscopy of
  molecular aggregates,'' J. Chem. Phys. {\bf 105}, 4565 (1996).

\bibitem{Tan939496}
Y.~Tanimura and S.~Mukamel, \newblock ``Two-dimensional femtosecond vibrational
  spectroscopy of liquids,'' J. Chem. Phys. {\bf 99}, 9496 (1993).

\bibitem{Tan943049}
Y.~Tanimura and S.~Mukamel, \newblock ``Multistate quantum Fokker-Planck
  approach to nonadiabatic wave packet dynamics in pump-probe spectroscopy,''
  J. Chem. Phys. {\bf 101}, 3049 (1994).

\bibitem{Dor132746}
K.~E. Dorfman, D.~V. Voronine, S.~Mukamel, and M.~O. Scully, \newblock
  ``Photosynthetic reaction center as a quantum heat engine,'' Proc. Natl.
  Acad. Sci. {\bf 110}, 2746 (2013).

\bibitem{Kun22015101}
S.~Kundu, R.~Dani, and N.~Makri, \newblock ``B800-to-B850 relaxation of
  excitation energy in bacterial light harvesting: All-state, all-mode path
  integral simulations,'' J. Chem. Phys. {\bf 157}, 015101 (2022).

\bibitem{Tan89101}
Y.~Tanimura and R.~Kubo, \newblock ``Time evolution of a quantum system in
  contact with a nearly Gaussian-Markovian noise bath,'' J. Phys. Soc. Jpn.
  {\bf 58}, 101 (1989).

\bibitem{Tan906676}
Y.~Tanimura, \newblock ``Nonperturbative expansion method for a quantum system
  coupled to a harmonic-oscillator bath,'' Phys. Rev. A {\bf 41}, 6676 (1990).

\bibitem{Tan06082001}
Y.~Tanimura, \newblock ``Stochastic Liouville, Langevin, Fokker-Planck, and
  master equation approaches to quantum dissipative systems,'' J. Phys. Soc.
  Jpn. {\bf 75}, 082001 (2006).

\bibitem{Yan04216}
Y.~A. Yan, F.~Yang, Y.~Liu, and J.~S. Shao, \newblock ``Hierarchical approach
  based on stochastic decoupling to dissipative systems,'' Chem. Phys. Lett.
  {\bf 395}, 216 (2004).

\bibitem{Xu05041103}
R.~X. Xu, P.~Cui, X.~Q. Li, Y.~Mo, and Y.~J. Yan, \newblock ``Exact quantum
  master equation via the calculus on path integrals,'' J. Chem. Phys. {\bf
  122}, 041103 (2005).

\bibitem{Xu07031107}
R.~X. Xu and Y.~J. Yan, \newblock ``Dynamics of quantum dissipation systems
  interacting with bosonic canonical bath: Hierarchical equations of motion
  approach,'' Phys. Rev. E {\bf 75}, 031107 (2007).

\bibitem{Tan20020901}
Y.~Tanimura, \newblock ``Numerically ``exact'' approach to open quantum
  dynamics: The hierarchical equations of motion (HEOM),'' J. Chem. Phys {\bf
  153}, 020901 (2020).

\bibitem{Sha045053}
J.~S. Shao, \newblock ``Decoupling quantum dissipation interaction via
  stochastic fields,'' J. Chem. Phys. {\bf 120}, 5053 (2004).

\bibitem{Hsi18014104}
C.-Y. Hsieh and J.~Cao, \newblock ``A unified stochastic formulation of
  dissipative quantum dynamics. II. Beyond linear response of spin baths,'' J.
  Chem. Phys. {\bf 148}, 014104 (2018).

\bibitem{Imr02}
I.~Imry,
\newblock {\em Introduction to Mesoscopic Physics},
\newblock Oxford university press, 2002.

\bibitem{Hau08}
H.~Haug and A.-P. Jauho,
\newblock {\em Quantum Kinetics in Transport and Optics of Semiconductors},
\newblock Springer-Verlag, Berlin, 2nd, substantially revised edition, 2008,
\newblock Springer Series in Solid-State Sciences 123.

\bibitem{Zha15024112}
H.~D. Zhang, R.~X. Xu, X.~Zheng, and Y.~J. Yan, \newblock ``Nonperturbative
  spin-boson and spin-spin dynamics and nonlinear Fano interferences: A unified
  dissipaton theory based study,'' J. Chem. Phys. {\bf 142}, 024112 (2015).

\bibitem{Du20034102}
P.~L. Du, Y.~Wang, R.~X. Xu, H.~D. Zhang, and Y.~J. Yan, \newblock
  ``System-bath entanglement theorem with Gaussian environments,'' J. Chem.
  Phys. {\bf 152}, 034102 (2020).

\bibitem{Che21244105}
Z.~H. Chen, Y.~Wang, R.~X. Xu, and Y.~J. Yan, \newblock ``Correlated
  vibration-solvent effects on the non-Condon exciton spectroscopy,'' J. Chem.
  Phys. {\bf 154}, 244105 (2021).

\bibitem{Mei922512}
Y.~Meir and N.~S. Wingreen, \newblock ``Landauer formula for the current
  through an interacting electron region,'' Phys. Rev. Lett. {\bf 68}, 2512
  (1992).

\bibitem{Gru1624514}
D.~Gruss, K.~A. Velizhanin, and M.~Zwolak, \newblock ``Landauer's formula with
  finite-time relaxation: Kramers' crossover in electronic transport,'' Phys.
  Rep. {\bf 6}, 24514 (2016).

\bibitem{Wan20041102}
Y.~Wang, R.~X. Xu, and Y.~J. Yan, \newblock ``Entangled system-and-environment
  dynamics: Phase-space dissipaton theory,'' J. Chem. Phys. {\bf 152}, 041102
  (2020).

\bibitem{Du212155}
P.~L. Du, Z.~H. Chen, Y.~Su, Y.~Wang, R.~X. Xu, and Y.~J. Yan, \newblock
  ``Nonequilibrium system--bath entanglement theorem versus heat transport,''
  Chem. J. Chin. Univ. {\bf 42}, 2155 (2021).

\bibitem{Wan22044102}
Y.~Wang, Z.~H. Chen, R.~X. Xu, X.~Zheng, and Y.~J. Yan, \newblock ``A
  statistical quasi--particles thermofield theory with Gaussian environments:
  System--bath entanglement theorem for nonequilibrium correlation functions,''
  J. Chem. Phys. {\bf 157}, 044012 (2022).

\bibitem{Kir35300}
J.~G. Kirkwood, \newblock ``Statistical mechanics of fluid mixtures,'' J. Chem.
  Phys. {\bf 3}, 300 (1935).

\bibitem{Gon20154111}
H.~Gong, Y.~Wang, H.~D. Zhang, Q.~Qiao, R.~X. Xu, X.~Zheng, and Y.~J. Yan,
  \newblock ``Equilibrium and transient thermodynamics: A unified
  dissipaton--space approach,'' J. Chem. Phys. {\bf 153}, 154111 (2020).

\bibitem{Gon20214115}
H.~Gong, Y.~Wang, H.~D. Zhang, R.~X. Xu, X.~Zheng, and Y.~J. Yan, \newblock
  ``Thermodynamic free--energy spectrum theory for open quantum systems,'' J.
  Chem. Phys. {\bf 153}, 214115 (2020).

\bibitem{Yan14054105}
Y.~J. Yan, \newblock ``Theory of open quantum systems with bath of electrons
  and phonons and spins: Many-dissipaton density matrixes approach,'' J. Chem.
  Phys. {\bf 140}, 054105 (2014).

\bibitem{Yan16110306}
Y.~J. Yan, J.~S. Jin, R.~X. Xu, and X.~Zheng, \newblock ``Dissipaton equation
  of motion approach to open quantum systems,'' Frontiers Phys. {\bf 11},
  110306 (2016).

\bibitem{Wan22170901}
Y.~Wang and Y.~J. Yan, \newblock ``Quantum mechanics of open systems:
  Dissipaton theories,'' J. Chem. Phys. {\bf 157}, 170901 (2022).

\bibitem{Xu151816}
R.~X. Xu, H.~D. Zhang, X.~Zheng, and Y.~J. Yan, \newblock ``Dissipaton equation
  of motion for system-and-bath interference dynamics,'' Sci. China Chem. {\bf
  58}, 1816 (2015),
\newblock Special Issue: Lemin Li Festschrift.

\bibitem{Zha16204109}
H.~D. Zhang, Q.~Qiao, R.~X. Xu, and Y.~J. Yan, \newblock ``Effects of
  Herzberg--Teller vibronic coupling on coherent excitation energy transfer,''
  J. Chem. Phys. {\bf 145}, 204109 (2016).

\bibitem{Xu17395}
R.~X. Xu, Y.~Liu, H.~D. Zhang, and Y.~J. Yan, \newblock ``Theory of quantum
  dissipation in a class of non-Gaussian environments,'' Chin. J. Chem. Phys.
  {\bf 30}, 395 (2017).

\bibitem{Xu18114103}
R.~X. Xu, Y.~Liu, H.~D. Zhang, and Y.~J. Yan, \newblock ``Theories of quantum
  dissipation and nonlinear coupling bath descriptors,'' J. Chem. Phys. {\bf
  148}, 114103 (2018).

\bibitem{Su224554}
Y.~Su, Z.~H. Chen, H.-J. Zhu, Y.~Wang, L.~Han, R.~X. Xu, and Y.~J. Yan,
  \newblock ``Electron transfer under the Floquet modulation in
  donor--bridge--acceptor systems,'' J. Phys. Chem. A {\bf 126}, 4554 (2022).

\bibitem{Gon22054109}
H.~Gong, Y.~Wang, X.~Zheng, R.~X. Xu, and Y.~J. Yan, \newblock ``Nonequilibrium
  work distributions in quantum impurity system--bath mixing processes,'' J.
  Chem. Phys. {\bf 157}, 054109 (2022).

\bibitem{Che21174111}
Z.~H. Chen, Y.~Wang, R.~X. Xu, and Y.~J. Yan, \newblock ``Quantum dissipation
  with nonlinear environment couplings: Stochastic fields dressed dissipaton
  equation of motion approach,'' J. Chem. Phys. {\bf 155}, 174111 (2021).

\bibitem{Zha18780}
H.~D. Zhang, R.~X. Xu, X.~Zheng, and Y.~J. Yan, \newblock ``Statistical
  quasi-particle theory for open quantum systems,'' Mol. Phys. {\bf 116}, 780
  (2018),
\newblock Special Issue, ``Molecular Physics in China''.

\bibitem{Hu10101106}
J.~Hu, R.~X. Xu, and Y.~J. Yan, \newblock ``Pad\'{e} spectrum decomposition of
  Fermi function and Bose function,'' J. Chem. Phys. {\bf 133}, 101106 (2010).

\bibitem{Hu11244106}
J.~Hu, M.~Luo, F.~Jiang, R.~X. Xu, and Y.~J. Yan, \newblock ``Pad\'{e} spectrum
  decompositions of quantum distribution functions and optimal hierarchial
  equations of motion construction for quantum open systems,'' J. Chem. Phys.
  {\bf 134}, 244106 (2011).

\bibitem{Din11164107}
J.~J. Ding, J.~Xu, J.~Hu, R.~X. Xu, and Y.~J. Yan, \newblock ``Optimized
  hierarchical equations of motion for Drude dissipation with applications to
  linear and nonlinear optical responses,'' J. Chem. Phys. {\bf 135}, 164107
  (2011).

\bibitem{Din12224103}
J.~J. Ding, R.~X. Xu, and Y.~J. Yan, \newblock ``Optimizing hierarchical
  equations of motion for quantum dissipation and quantifying quantum bath
  effects on quantum transfer mechanisms,'' J. Chem. Phys. {\bf 136}, 224103
  (2012).

\bibitem{Zhe121129}
X.~Zheng, R.~X. Xu, J.~Xu, J.~S. Jin, J.~Hu, and Y.~J. Yan, \newblock
  ``Hierarchical equations of motion for quantum dissipation and quantum
  transport,'' Prog. Chem. {\bf 24}, 1129 (2012).

\bibitem{Che22221102}
Z.~H. Chen, Y.~Wang, X.~Zheng, R.~X. Xu, and Y.~J. Yan, \newblock ``Universal
  time-domain Prony fitting decomposition for optimized hierarchical quantum
  master equations,'' J. Chem. Phys. {\bf 156}, 221102 (2022).

\bibitem{Cui19024110}
L.~Cui, H.~D. Zhang, X.~Zheng, R.~X. Xu, and Y.~J. Yan, \newblock ``Highly
  efficient and accurate sum--over--poles expansion of Fermi and Bose functions
  at near zero temperatures: Fano spectrum decomposition scheme,'' J. Chem.
  Phys. {\bf 151}, 024110 (2019).

\bibitem{Zha20064107}
H.~D. Zhang, L.~Cui, H.~Gong, R.~X. Xu, X.~Zheng, and Y.~J. Yan, \newblock
  ``Hierarchical equations of motion method based on Fano spectrum
  decomposition for low temperature environments,'' J. Chem. Phys. {\bf 152},
  064107 (2020).

\bibitem{Zho08034106}
Y.~Zhou and J.~S. Shao, \newblock ``Solving the spin-boson model of strong
  dissipation with flexible random-deterministic scheme,'' J. Chem. Phys. {\bf
  128}, 034106 (2008).

\bibitem{Liu14134106}
H.~Liu, L.~L. Zhu, S.~M. Bai, and Q.~Shi, \newblock ``Reduced quantum dynamics
  with arbitrary bath spectral densities: Hierarchical equations of motion
  based on several different bath decomposition schemes,'' J. Chem. Phys. {\bf
  140}, 134106 (2014).

\bibitem{Tan15224112}
Z.~F. Tang, X.~L. Ouyang, Z.~H. Gong, H.~B. Wang, and J.~L. Wu, \newblock
  ``Extended hierarchy equation of motion for the spin-boson model,'' J. Chem.
  Phys. {\bf 143}, 224112 (2015).

\bibitem{Nak18012109}
K.~Nakamura and Y.~Tanimura, \newblock ``Hierarchical Schr\"odinger equations
  of motion for open quantum dynamics,'' Phys. Rev. A {\bf 98}, 012109 (2018).

\bibitem{Ike20204101}
T.~Ikeda and G.~D. Scholes, \newblock ``Generalization of the hierarchical
  equations of motion theory for efficient calculations with arbitrary
  correlation functions,'' J. Chem. Phys. {\bf 152}, 204101 (2020).

\bibitem{Lam193721}
N.~Lambert, S.~Ahmed, M.~Cirio, and F.~Nori, \newblock ``Modelling the
  ultra-strongly coupled spin-boson model with unphysical modes,'' Nature Comm.
  {\bf 10}, 3721 (2019).

\bibitem{Fey63118}
R.~P. Feynman and F.~L. \mbox{Vernon, Jr.}, \newblock ``The theory of a general
  quantum system interacting with a linear dissipative system,'' Ann. Phys.
  {\bf 24}, 118 (1963).

\bibitem{Zhu12194106}
L.~L. Zhu, H.~Liu, W.~W. Xie, and Q.~Shi, \newblock ``Explicit system-bath
  correlation calculated using the hierarchical equations of motion method,''
  J. Chem. Phys. {\bf 137}, 194106 (2012).

\bibitem{Dal01053813}
B.~J. Dalton, S.~M. Barnett, and B.~M. Garraway, \newblock ``Theory of
  pseudomodes in quantum optical processes,'' Phys. Rev. A {\bf 64}, 053813
  (2001).

\bibitem{Tam18030402}
D.~Tamascelli, A.~Smirne, S.~F. Huelga, and M.~B. Plenio, \newblock
  ``Nonperturbative treatment of non-Markovian dynamics of open quantum
  systems,'' Phys. Rev. Lett. {\bf 120}, 030402 (2018).

\bibitem{Tam19090402}
D.~Tamascelli, A.~Smirne, J.~Lim, S.~F. Huelga, and M.~B. Plenio, \newblock
  ``Efficient simulation of finite-temperature open quantum systems,'' Phys.
  Rev. Lett. {\bf 123}, 090402 (2019).

\bibitem{Che19123035}
F.~Chen, E.~Arrigoni, and M.~Galperin, \newblock ``Markovian treatment of
  non-Markovian dynamics of open Fermionic systems,'' New J. Phys. {\bf 21},
  123035 (2019).

\bibitem{Hu907078}
H.~Gang and H.~Haken, \newblock ``Steepest-descent approximation of stationary
  probability distribution of systems driven by weak colored noise,'' Phys.
  Rev. A {\bf 41}, 7078 (1990).

\bibitem{Han0611438}
P.~Han, R.~X. Xu, B.~Q. Li, J.~Xu, P.~Cui, Y.~Mo, and Y.~J. Yan, \newblock
  ``Kinetics and thermodynamics of electron transfer in Debye solvents: An
  analytical and nonperturbative reduced density matrix theory,'' J. Phys.
  Chem. B {\bf 110}, 11438 (2006).

\bibitem{Che07438}
Y.~Chen, R.~X. Xu, H.~W. Ke, and Y.~J. Yan, \newblock ``Electron transfer
  theory revisit: Motional narrowing induced non-Markovian rate processes,''
  Chin. J. Chem. Phys. {\bf 20}, 438 (2007).

\bibitem{Sha15}
R.~Shanmugam and R.~Chattamvelli, \newblock ``Skewness and Kurtosis,'' in {\em
  Statistics for Scientists and Engineers}, chapter~4, pages 89--110, John
  Wiley \& Sons, Ltd, 2015.

\bibitem{Shi09164518}
Q.~Shi, L.~P. Chen, G.~J. Nan, R.~X. Xu, and Y.~J. Yan, \newblock ``Electron
  transfer dynamics: Zusman equation versus exact theory,'' J. Chem. Phys. {\bf
  130}, 164518 (2009).

\bibitem{Gu851310}
Y.~Gu, \newblock ``Group-theoretical formalism of quantum mechanics based on
  quantum generalization of characteristic functions,'' Phys. Rev. A {\bf 32},
  1310 (1985).

\end{thebibliography}

\end{document}